\documentclass[aps,prl,twocolumn,longbibliography,groupedaddress,showpacs,floatfix,superscriptaddress]{revtex4-1}
\usepackage{graphicx}  
\usepackage{dcolumn}   
\usepackage[english]{babel}
\usepackage{bm}        
\usepackage{amsmath, amsthm, amssymb}
\usepackage{bbold} 
\usepackage{mathrsfs}
\usepackage{array}
\usepackage[usenames]{color}
\usepackage{color}
\usepackage{soul} 
\usepackage{ulem} 
\usepackage{filecontents}

\emergencystretch=\maxdimen
\begin{document}

\title{Magnetism and charge order in the honeycomb lattice}
\author{Natanael C. Costa} 
\email{natanael@if.ufrj.br}
\email{ndecarva@sissa.it}
\affiliation{International School for Advanced Studies (SISSA),
Via Bonomea 265, 34136, Trieste, Italy}
\affiliation{Instituto de F\'isica, Universidade Federal do Rio de Janeiro
Cx.P. 68.528, 21941-972 Rio de Janeiro RJ, Brazil}
\author{Kazuhiro Seki}
\affiliation{
  Computational Quantum Matter Research Team, RIKEN, Center for Emergent Matter Science (CEMS), Saitama 351-0198, Japan}
\author{Sandro Sorella}
\affiliation{International School for Advanced Studies (SISSA),
Via Bonomea 265, 34136, Trieste, Italy}

\begin{abstract}
Despite being relevant to better understand the properties of honeycomb-like systems, as graphene-based compounds, the electron-phonon interaction is commonly disregarded in theoretical approaches.
That is, the effects of phonon fields on \textit{interacting} Dirac electrons is an open issue, in particular when investigating long-range ordering.
Thus, here we perform unbiased quantum Monte Carlo simulations to examine the Hubbard-Holstein model (HHM) in the half-filled honeycomb lattice.
By performing careful finite-size scaling analysis, we identify semimetal-to-insulator quantum critical points, and determine the behavior of the antiferromagnetic and charge-density wave phase transitions.
We have, therefore, established the ground state phase diagram of the HHM for intermediate interaction strength, determining its behavior for different phonon frequencies.
Our findings represent a complete description of the model, and may shed light on the emergence of many-body properties in honeycomb-like systems.
\end{abstract}


\maketitle

\noindent
\underbar{Introduction:}
The electronic properties of quasi-two-dimensional materials have been under intense debate over the past years, due to the emergence of a plethora of many-body phenomena\,\cite{Kotov12,Bhimanapati15,Manzeli17,Chen16}.
For instance, the quasi-2D transition-metal dichalcogenides may exhibit charge-density wave (CDW), superconductivity (SC), and also topological properties, with the electron-phonon (\textit{e-ph}) interaction being the key ingredient\,\cite{Manzeli17,Chen16,Rossnagel11,Joe14}.
Another important material, and probably the most prominent 2D one, is graphene, whose electronic dispersion leads to Dirac-like electrons\,\cite{CastroNeto09}.
Despite being considered a weakly interacting compound, electron correlation may play an important role in  graphene\,\cite{Kotov12}, as the occurrence of a semimetal-to-insulator transition driven by strain\,\cite{Si16,Lee12,Tang15,Chen18b,Sorella18}, and the remarkable emergence of SC in the twisted bilayer graphene\cite{Cao18a,Cao18b,Yankowitz19,Balents20}.
Moreover, these features have intensified the discussions about the relevant role played by the \textit{e-ph} coupling\,\cite{Eliel18,Wu18,Lian19,Angeli19}.
That is, to further understand the nature of these materials, it is of paramount importance to take into account \textit{both} \textit{e-e} and \textit{e-ph} interactions.
 
For the particular case of graphene, some insights about its metallic behavior have emerged from recent \textit{ab-initio} computational studies, aiming to determine the on-site \textit{e-e} interaction\,\cite{Wehling11,Zheng18a,Zheng18b}.
In fact, its on-site (Hubbard-like) interaction is finite, but smaller than the theoretical semimetal-to-insulator quantum critical point (QCP), $U/t\approx 3.8$, predicted by quantum Monte Carlo (QMC) simulations for the Hubbard model in the honeycomb lattice\,\cite{Assaad13,Otsuka16}.
One should notice that such mappings into effective Hamiltonians may underestimate the on-site \textit{e-e} interaction if the \textit{e-ph} coupling is not considered.
However, and interestingly, just a few theoretical studies have considered phonon effects on Dirac fermions\,\cite{Tse07,Stauber08,Classen14,Zhang19,Chen19,Feng20,Zhang20}, despite the relevance of (optical) phonon modes for graphene\,\cite{Piscanec07,Lazzeri08,Attaccalite10,Haberer13}.
Indeed, to the best of our knowledge, no exact results are available for the behavior of the antiferromagnetic (AFM) phase transition in the presence of phonon fields, on the honeycomb lattice.

\begin{figure}[t]
\centering
\includegraphics[scale=0.25]{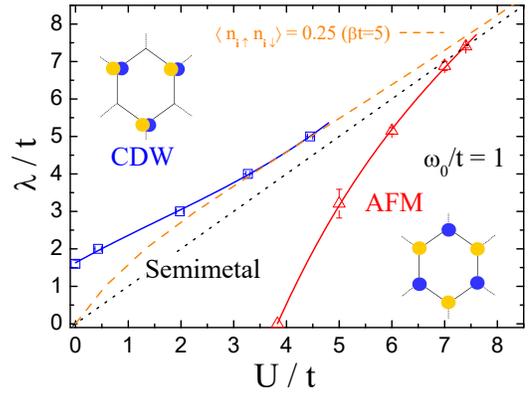}
\caption{The ground state phase diagram of the Hubbard-Holstein model in the half-filled honeycomb lattice, with fixed $\omega_{0}/t$.
  The dashed (orange) curve determines
  the parameter for which $\langle n_{\mathbf{i} \uparrow} n_{\mathbf{i} \downarrow} \rangle=0.25$ at a fixed finite temperature, while the continuous ones are just guides to the eyes.
Here, and in all subsequent figures, when not shown, error bars are smaller than symbol size.}
\label{fig:phase_diagram} 
\end{figure}

In this Letter, we fill this gap by investigating the properties of the Hubbard-Holstein model (HHM) in the half-filled honeycomb lattice.
The HHM describes interacting (itinerant) electrons locally coupled to dispersionless phonon degrees of freedom, treating the \textit{e-e} and \textit{e-ph} interactions on equal footing.
While the Coulomb repulsion leads to spin fluctuations, favoring an AFM phase, the \textit{e-ph} coupling favors CDW (and/or SC); an interplay/competition between these phases is therefore expected.
In view of this, here we examine the charge and magnetic properties of the HHM, performing unbiased QMC simulations for lattices up to 648 sites (linear size $L=18$).
The ground state phase diagram, displayed in Fig.\,\ref{fig:phase_diagram}, highlights some of our main findings:
[i] we present precise semimetal-to-CDW/AFM QCPs, which
[ii] are quite sensitive on the choice of external parameters;
[iii] in the limit case of equal \textit{e-e} and \textit{e-ph} couplings ($U=\lambda$), spin fluctuations dominate, and an AFM transition occurs for a finite interaction strength.
In addition, [iv] we also present a complete description of the behavior of these QCPs as the phonon frequency is varied, comparing our findings with other similar models in literature.


\noindent
\underbar{Model and Methodologies:}
The Hubbard-Holstein Hamiltonian reads
\begin{align} \label{eq:HHM_hamil}
\nonumber \mathcal{H} = &
-t \sum_{\langle \mathbf{i}, \mathbf{j} \rangle, \sigma} 
\big(d^{\dagger}_{\mathbf{i} \sigma} d^{\phantom{\dagger}}_{\mathbf{j} \sigma} + {\rm h.c.} \big)
- \mu \sum_{\mathbf{i}, \sigma} n^{\phantom{\dagger}}_{\mathbf{i}, \sigma}
+ U \sum_{\mathbf{i}} n_{\mathbf{i}\uparrow} n_{\mathbf{i}\downarrow}
\\
&
+ \sum_{ \mathbf{i} }
\bigg( \frac{\hat{P}^{2}_{\mathbf{i}}}{2 M} + \frac{M \omega^{2}_{0} }{2} \hat{X}^{2}_{\mathbf{i}}\bigg)
- g \sum_{\mathbf{i}, \sigma} n_{\mathbf{i}\sigma} \hat{X}_{\mathbf{i}}
,
\end{align}
with $d^{\dagger}_{\mathbf{i} \sigma}$ ($d^{\phantom{\dagger}}_{\mathbf{i} \sigma}$)
being a creation (annihilation) operator for electrons with spin $\sigma\,(=\uparrow,\downarrow)$
at a given site $\mathbf{i}$.
Here, the sums run over a 2D honeycomb lattice under periodic boundary conditions, with $\langle \mathbf{i}, \mathbf{j} \rangle$ denoting nearest neighbors. 
The first two terms on the right hand side of Eq.\,\eqref{eq:HHM_hamil} correspond to the hopping of electrons, and their chemical potential ($\mu$) term, respectively, with $n^{\phantom{\dagger}}_{\mathbf{i}\sigma}\equiv d^{\dagger}_{\mathbf{i} \sigma} d^{\phantom{\dagger}}_{\mathbf{i} \sigma}$ being site number operators.
The on-site Coulomb repulsion between electrons is given by the third term.
The phonon modes are added by quantum harmonic oscillators with frequency $\omega_{0}$ on each site of the lattice (fourth term), while the local \textit{e-ph} interaction is described in the last term.
Hereafter, we define the mass of the harmonic oscillators ($M$) and the lattice constant as unity, and set the energy scale in units of the hopping integral $t$.

In order to facilitate the discussions throughout the paper, it is important to remark that the \textit{e-ph} coupling leads to polaron formation, and to retardation effects in the {\it e-e} interactions.
Therefore, one should define additional parameters to take into account these energy scales, which are conveniently given by perturbation theory~\cite{Berger95}.
Within this approach, (i) the energy scale for polarons -- which also determines the effective attractive interaction between electrons -- is $\lambda \equiv g^{2} / \omega^{2}_{0}$, while (ii) the adiabaticity ratio (for retardation effects) is $\omega_{0}/t$.
In addition, we also define $U_{\rm eff} \equiv U - \lambda$, a parameter that gives insights about the on-site {\it e-e} interactions, what is also relevant for our QMC methodology, as discussed below.

We examine the properties of Eq.\,\eqref{eq:HHM_hamil} by combining the analyses of the projective ground state auxiliary-field QMC (AFQMC)~\cite{Sorella89,Blankenbecler81,becca17}, and the finite temperature determinant QMC (DQMC) methods~\cite{Blankenbecler81,Hirsch83,Hirsch85,Santos03,gubernatis16}.
Both methodologies may suffer from the infamous minus-sign problem, depending on the electron filling, temperature (projection time), or the kind/strength of electronic interactions\,\cite{Loh90}.
Despite this, a sign-free AFQMC approach may be employed for the half-filling HHM, when keeping $U \geq \lambda$ (i.e., $U_{\rm eff} \geq 0$), as described in Ref.\,\onlinecite{Karakuzu18}.
We use this approach for our AFQMC method, improving it by the implementation of an inversion sampling algorithm; see, e.g., the Supplementary Materials (SM). 
The complementary region $U < \lambda$ is investigated by means of DQMC simulations\,\cite{Johnston13}, for values of interaction strengths in which the average sign is high enough to obtain precise results -- see also the discussions in the SM.
Thus, using the AFQMC and DQMC methods complementarily, we are able to investigate the half-filling phase diagram of the HHM, and probe the existence of long-range ordered phases.

In particular, we examine the charge and magnetic properties of the HHM, which is accomplished by measuring the CDW and AFM structure factors:
$S_{\rm cdw}(\mathbf{q}) = \frac{1}{N} \sum_{\mathbf{r}_{i}, \mathbf{r}_{j}} e^{-{\rm i}\mathbf{q}\cdot(\mathbf{r}_{i} - \mathbf{r}_{j})} \langle (n_{A,\mathbf{r}_{i}} - n_{B,\mathbf{r}_{i}}) (n_{A,\mathbf{r}_{j}} - n_{B,\mathbf{r}_{j}}) \rangle $, and 
$S_{\rm afm}(\mathbf{q}) = \frac{1}{N} \sum_{\mathbf{r}_{i}, \mathbf{r}_{j}} e^{-{\rm i}\mathbf{q}\cdot(\mathbf{r}_{i} - \mathbf{r}_{j})} \langle (S^{z}_{A,\mathbf{r}_{i}} - S^{z}_{B,\mathbf{r}_{i}}) (S^{z}_{A,\mathbf{r}_{j}} - S^{z}_{B,\mathbf{r}_{j}}) \rangle $,
with
$n_{\alpha, \mathbf{r}_{i}}=n_{\alpha, \mathbf{r}_{i} \uparrow} + n_{\alpha, \mathbf{r}_{i} \downarrow}$, and
$S^{z}_{\alpha, \mathbf{r}_{i}}=\frac{1}{2}(n_{\alpha, \mathbf{r}_{i} \uparrow} - n_{\alpha, \mathbf{r}_{i} \downarrow}$).
Here, $\alpha =A, B$ labels the sublattices, while $\mathbf{r}_{i}$ is the unit cell position of a given site $\mathbf{i}$, with $N= 2 L \times L$ being the total number of lattice sites.
Given this, one may probe the AFM/CDW critical behavior by means of the correlation ratio
\begin{align}\label{eq:Rc}
R_{\nu}(L) = 1 - \frac{S_{\nu}(\mathbf{q}+\delta\mathbf{q})}{S_{\nu}(\mathbf{q})},
\end{align}
with $\mathbf{q}=(0,0)$ and $\mathbf{q} + \delta\mathbf{q}$ its neighboring wavevector,
and $\nu$ labelling `cdw' or `afm'.
The critical points are estimated by a finite-size scaling (FSS) analysis of the crossing points of $R_{\nu}(L)$ for different lattice sizes~\cite{Kaul15,Sato18,Liu18,Darmawan18}.
We define $\beta \propto L$ (or $\tau \propto L$), with $\beta$ being the inverse of temperature for the DQMC method (and $\tau$ the projection time in the AFQMC one) assuming the Lorentz invariance at QCPs\,\cite{Assaad13,Otsuka16,Toldin15}.
In what follows, we set $\langle n_{\mathbf{i}\sigma} \rangle = 1/2$.

\begin{figure}[t]
\centering
\includegraphics[scale=0.30]{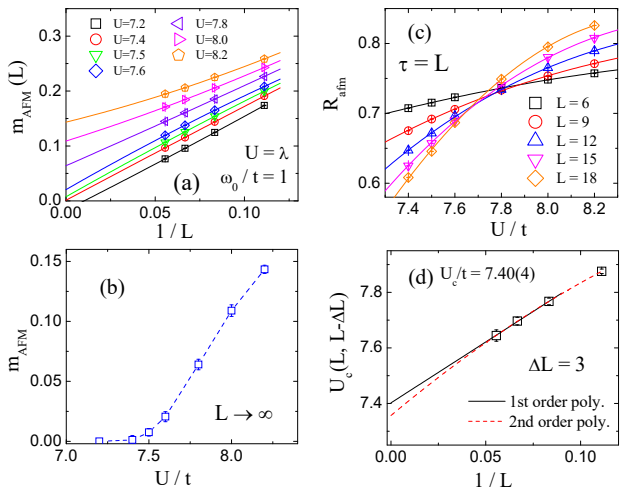}
\caption{Projective AFQMC results for the antiferromagnetic response along the line $U=\lambda$, for fixed $\omega_{0}/t=1$.
  (a) The linear order parameter as a function of the system size $L$, for different $U/t$, and (b) its extrapolated values to $L\to\infty$.
(c) The AFM correlation ratio as a function of $U/t$, and (d) its the crossing points for two consecutive lattice sizes.
} 
\label{fig1:AFM1} 
\end{figure}

\noindent
\underbar{Results:}
Let us first consider the most challenging case: the ground state behavior for $U=\lambda$.
For this choice of parameters, electronic and phononic interactions are equal, and the occurrence of long-range order is less evident.
Our AFQMC results give a low response for the charge-charge correlation functions for any interaction strength (not shown), while there is an enhancement for the spin-spin ones. 
The possible occurrence of long-range magnetic order at ground state may be given by the linear order parameter
$m_{\rm afm}(L) = \frac{1}{N} \sum_{\mathbf{r}_{i}} \langle S^{z}_{A,\mathbf{r}_{i}} - S^{z}_{B,\mathbf{r}_{i}} \rangle$,
and its FSS analysis to the thermodynamic limit~\footnote{Here we call the attention of the reader that, in the AFQMC method, the spin symmetry is broken along the $z$-direction for the trial wavefunction.}.
Figure \ref{fig1:AFM1}\,(a) displays $m_{\rm afm}(L)$ for different values of $U/t$ and lattice sizes, fixing $\omega_{0}/t=1$.
The solid lines are polynomial extrapolations to $L\to\infty$, whose values are used to plot Fig.\,\ref{fig1:AFM1}\,(b).
Indeed, there is evidence for an AFM quantum phase transition at $U_{c}^{\rm afm}= \lambda_{c}^{\rm afm} \approx 7.4 t$.
It is worth noticing that the strength of the coupling needed for AFM transition is significantly increased by the
{\it e-ph} coupling as compared to that $U_c^{\rm afm}/t=3.85(2)$ for the pure Hubbard model\,\cite{Assaad13,Otsuka16}.
We discuss such phononic effects in more detail later.

A thorough determination of this QCP may be obtained from the analysis of the correlation ratio, Eq.\,\eqref{eq:Rc}. 
Figure \ref{fig1:AFM1}\,(c) presents $R_{\rm afm}(L)$ as a function of $U/t$, and for the same parameters of panel (a).
The correlation ratio exhibits an enhancement for large $U/t$, which corresponds to an increase in the
$S_{\rm afm}(\mathbf{q}=0)$,
and is also suggestive of long-range order.
The crossing points between $R_{\rm afm}(L)$ for two consecutive lattice sizes -- defined here as $U_{c}(L,L-\Delta L)$ -- identify the critical region.
Figure \ref{fig1:AFM1}\,(d) displays these points, and their FSS analysis, which determines the AFM critical point as $U_{c}^{\rm afm}= \lambda_{c}^{\rm afm} \approx 7.40(4) t$, in good agreement with the analysis of the linear order parameter.
Therefore, hereafter our investigation of the critical points will be based the behavior of $R_{\nu}(L)$.

\begin{figure}[t]
\centering
\includegraphics[scale=0.28]{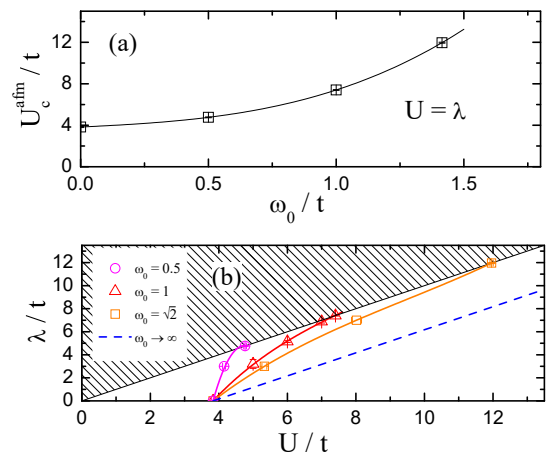}
\caption{The AFM critical points (a) along the line $U=\lambda$, and (b) for the general case of $U\geq\lambda$, for different phonon frequencies.
The dashed line corresponds to the AFM transition at $\omega_{0}\to\infty$ limit, while the hatched region indicates the forbidden domain for the employed AFQMC method.
}
\label{fig2:AFM2} 
\end{figure}

It is natural to seek how susceptible is this QCP to changes in the phonon frequencies.
Naively, for $U=\lambda$,
one would expect $U_{c}^{\rm afm} \to \infty$ as $\omega_{0}/t \to \infty$, due to the instantaneous interaction in the antiadiabatic limit.
To quantify this, we have repeated the above analysis for other values of $\omega_{0}/t$, the results of which are presented in Fig.\,\ref{fig2:AFM2}\,(a).
Notice that the QCP increases quickly as a function of $\omega_{0}/t$, reaching a quite strong $U_{c}^{\rm afm} = 11.96(5) t$ for $\omega_{0}/t=\sqrt{2}$.
In a direct comparison with the square lattice\,\cite{Costa20}, the honeycomb critical $U_{c}^{\rm afm}(\omega_{0})$ along the line $U=\lambda$ is way more susceptible to changes in $\omega_{0}/t$.
Additionally, for $\omega_{0} \to 0$, the QCP seems to approach the pure Hubbard case, $U_{c}^{\rm afm} = 3.85(2) t$.

We proceed to examine the AFM behavior for $U>\lambda$.
Here, one may analyze the behavior of $R_{\rm afm}(L)$ by fixing $\lambda$ while varying $U$, or vice versa.
Figure \ref{fig2:AFM2}\,(b) displays the AFM critical points for different $\omega_{0}/t$ and interaction strengths.
Notice that for small $\omega_{0}/t$ the transition line is sharp, with $U_{c}^{\rm afm}$ being very close to the pure Hubbard model case.
As the phonon frequency increases, the semimetal region grows, with substantial changes for $U_{c}^{\rm afm}$, in particular for larger $\lambda$.
Further increase of $\omega_{0}/t$ leads the critical points to gradually approach the AFM transition line for $\omega_{0} \to \infty$ case, depicted as the (blue) dashed line in Fig.\,\ref{fig2:AFM2}\,(b).
The latter is just a shift in the QCP of pure Hubbard model, keeping $U-\lambda=3.85$.
These results reveal how susceptible are the AFM critical points in the honeycomb lattice when \textit{e-ph} interactions are taken into account.

\begin{figure}[t]
\centering
\includegraphics[scale=0.31]{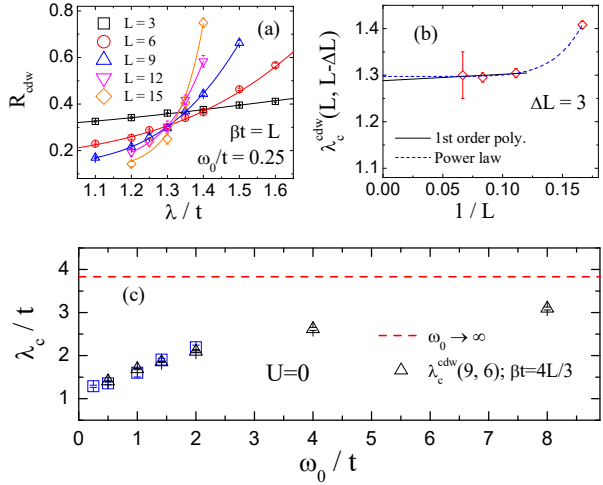}
\caption{Finite temperature DQMC results for (a) the charge correlation ratio as a function of $\lambda/t$, and fixed $\omega_{0}/t=0.25$ and $U=0$.
(b) The crossing points between $R_{\rm cdw}$ of two consecutive lattice sizes.
  (c) The CDW critical points for as a function of $\omega_{0}/t$ (blue squares symbols), and crossing points of $R_{\rm cdw}$ for $L=9$, and 6 (black triangle symbols); the dashed line is the critical point in the $\omega_{0} \to \infty$ limit.
When not shown, error bars are smaller than symbol size.}
\label{fig3:holstein} 
\end{figure}

We now turn to discuss the $\lambda > U$ case by DQMC simulations,
starting with the specific case of the pure Holstein model ($U=0$)~\footnote{For $U=0$ the DQMC approach is sign-problem-free.}, in which a CDW phase is expected for a finite $\lambda/t$~\cite{Zhang19,Chen19}.
Similarly to the previous analyses, here we determine the critical \textit{e-ph} coupling by investigating the crossings of $R_{\rm cdw}(L)$ for different system sizes, as displayed in Fig.\,\ref{fig3:holstein}\,(a), for fixed $\omega_{0}/t=0.25$.
The FSS analysis of $\lambda^{\rm cdw}_{c}(L,L-\Delta L)$ -- the crossing points between $R_{\rm cdw}(L)$ for consecutive lattice sizes -- leads to $\lambda^{\rm cdw}_{c}/t=1.29(3)$, as displayed in Fig.\,\ref{fig3:holstein}\,(b), confirming our expectation for the emergence of a charge ordered phase at ground state. 

It is worth examining the dependence of $\lambda^{\rm cdw}_{c}$ when $\omega_{0}/t$ is varied.
This analysis is displayed in Fig.\,\ref{fig3:holstein}\,(c), with the crossover from adiabatic to antiadiabatic limits.
Here, the blue square symbols are obtained from the FSS analysis of $\lambda^{\rm cdw}_{c}(L,L-\Delta L)$, while the black triangle symbols are just $\lambda^{\rm cdw}_{c}(9,6)$.
Notice that $\lambda^{\rm cdw}_{c}$ seems to have a
finite
value ($\approx 1.2t$) when $\omega_{0}\to 0$, while increasing monotonically and asymptotically to the attractive Hubbard model QCP for $\omega_{0}\to\infty$, i.e.~$\lambda^{\rm cdw (sc)}_{c}=3.8$, due a mapping between these models in this limit.
In addition, these results suggest that the phonon energy scale for the true antiadiabatic response should occur at quite large values of $\omega_{0}/t$\,\cite{Feng20,Xiao19b}, contrasting the common sense view about this issue.

\begin{figure}[t]
\centering
\includegraphics[scale=0.29]{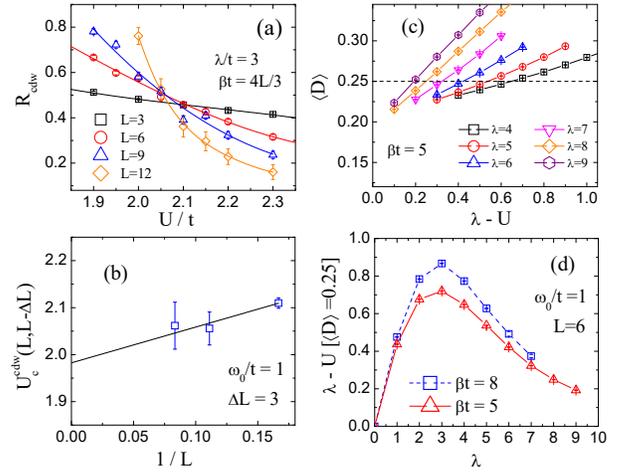}
\caption{Finite temperature DQMC results for (a) the charge correlation ratio as a function of $U/t$, for fixed $\lambda/t=3$ and $\omega_{0}/t=1$, and (b) its crossing points for two consecutive lattice sizes.
(c) The average of the double occupation as a function of $|U_{\rm eff}|$, for several values of $\lambda/t$, and fixed $\beta t=5$.
(d) The value of $|U_{\rm eff}|$ which gives $\langle D \rangle = 0.25$, as a function of $\lambda/t$, and fixed $\beta t=5$ and 8.
}
\label{fig4:CDW} 
\end{figure}

Finally, we analyze the general case for $\lambda > U$ ($U \neq 0$).
Since $U$ favors singly occupied sites, the CDW phase should be destroyed when \textit{e-e} interactions are turned on, as displayed in Fig.\,\ref{fig4:CDW}\,(a) for the behavior of $R_{\rm cdw}(L)$, with fixed $\lambda /t =3$, and $\omega_{0}/t=1$.
The crossing points of $R_{\rm cdw}(L)$ are displayed in Fig.\,\ref{fig4:CDW}\,(b), whose extrapolation yields $U^{\rm cdw}_{c}/t=1.98(7)$.
Other charge QCPs are obtained similarly, and presented in the phase diagram of Fig.\,\ref{fig:phase_diagram}.
It is important to mention that, although the semimetal-to-insulator phase transitions investigated in this Letter seem to be continuous within the range of parameters analyzed, insulator-to-insulator (AFM-to-CDW) first-order transitions could appear for $U \approx \lambda \gg t$ and small $\omega_0/t$, as observed in 1D chains\,\cite{Clay05,Tezuka07,Fehske08} and 2D square lattice\,\cite{Ohgoe17,Karakuzu17}.
The analysis of this regime is beyond the scope of this work; in particular, the projection on the ground state is challenging by DQMC simulations for large interaction strengths, due to the very low average sign for the product of determinants.

In spite of this, insights about the CDW phase at strong interaction strengths may be given by the behavior of charge-charge correlation functions at intermediate temperatures.
For instance, double occupation $\langle D\rangle = \langle n_{\mathbf{i} \uparrow} n_{\mathbf{i} \downarrow} \rangle$ is considerably favored for large $\lambda/t$ (and $U_{\rm eff}<0$), due to bipolaron formation~\cite{Freericks96,Han20}, thus can be employed to signal a change in this regime.
Therefore, we present in Fig.\,\ref{fig4:CDW}\,(c) the behavior of $\langle D\rangle$ as a function of $|U_{\rm eff}|$, for fixed $\beta/t=5$
\footnote{Since the double occupation is a local quantity, it is less affected by the system size, and $L=6$ seems a reasonable size for this case.}. 
Notice that the slope of the curves increases gradually with $\lambda/t$, corresponding to a quicker change from a singly to a doubly occupied site, features typical from AFM and CDW phases, respectively.
A rough manner to identify the critical region (in the strong coupling limit) is determining the $U_{\rm eff}$ for which $\langle D\rangle = 0.25$, as presented in Fig.\,\ref{fig4:CDW}\,(d) for $\beta/t=5$ and 8; being equivalent to the noninteracting case.
Indeed, these curves for $\lambda/t \gtrsim 3$ are very close to the charge QCPs presented in Fig.\,\ref{fig:phase_diagram}, and we expect that the $\lambda_{c}^{\rm cdw}$ for larger couplings should be around these curves.


\noindent
\underbar{Conclusions:}
In this Letter, we have presented a complete picture of the effects of \textit{e-ph} coupling on interacting Dirac electrons.
In summary, we examined the occurrence of semimetal-to-AFM/CDW quantum phase transitions for the HHM in the half-filled honeycomb lattice, using unbiased AFQMC and DQMC methods.
From a rigorous FSS analysis of the correlation ratio, we obtained precise QCPs making up the ground state phase diagram of the model for weak/intermediate interaction strengths.
Moreover, we have also presented a quantitative description of the effects of phonon frequency on the AFM/CDW phases, from the adiabatic to antiadiabatic limits.
In particular, we noticed that the AFM critical region is very susceptible to changes in $\omega_{0}/t$, when the on-site \textit{e-e} and the \textit{e-ph} interactions are on the same order of magnitude.

As a final remark, it is important comparing our HHM results with those of alternative models.
For instance, the phase diagram of the extended Hubbard model in the honeycomb lattice\,\cite{Herbut06,Wu14,Schuler18,Shao20} is qualitatively similar to our Fig.\,\ref{fig:phase_diagram}.
However, the intrinsic long-range character of phonon-induced interactions, roughly controlled by the adiabaticity ratio, leads to substantial changes in the AFM/CDW phase boundaries, a feature nowhere found in the extended Hubbard case.
As a step towards this end, further neighbor \textit{e-e} interactions are demanded, as accomplished by a Lang-Firsov transformation~\cite{Wang19}, but being a challenge for QMC simulations~\cite{Hohenadler04}.

Under some circumstances, adding long-range Coulomb interaction is feasible by QMC methodologies\,\cite{Tang18}, and it leads to an AFM phase diagram very similar to ours.
Nonetheless, we emphasize that in the HHM the pattern of the interactions may become unusual, i.e.~varying from repulsive to attractive depending on the phonon modes\,\cite{Wang19,Costa18}.
This property, without a direct analog in the Coulomb case, has been suggested as being relevant for the enhancement of pairing\,\cite{Wang19}.
In view of this, the inclusion of phonon fields seems fundamental to further understand the many-body nature of honeycomb-like compounds, and we expect that our findings shed light on it.

\begin{acknowledgments}
\underbar{Acknowledgements:}
Computational resources were provided by CINECA supercomputer (PRACE-2019204934).
S.S.~and N.C.C.~acknowledge PRACE for awarding them access to Marconi at CINECA, Italy.
S.S.~also acknowledges financial support from PRIN 2017BZPKSZ.
N.C.C.~thank R.R.\,dos Santos for the useful discussions, and acknowledges the Brazilian Agencies CAPES and CNPq for partially funding this project.
K.S.~is supported by Grant-in-Aid for Research Activity start-up (No.~JP19K23433).  

\end{acknowledgments}


\bibliography{bibHH03}

\clearpage
\noindent
{\Large Supplementary Material for: Magnetism and charge order in the honeycomb lattice}

\makeatletter
\renewcommand{\figurename}{Supplementary Figure}
\renewcommand{\tablename}{Supplementary Table}
\renewcommand{\theequation}{S\@arabic\c@equation}
\setcounter{equation}{0}
\setcounter{figure}{0}
\makeatletter

\subsection*{Supplementary Notes}


\noindent
\underbar{Inversion sampling in the AFQMC method}:
The important property that we are going to use here is that, within 
the  continuous Hubbard-Stratonovich (HS) transformation, 
it is possible  to compute explicitly   
the ratio of the weight $W(\{ \sigma \})$ when a single HS variable is changed, at a  given lattice point $i$ and time slice $l$, $\delta \sigma \equiv \sigma_{il} \to \sigma^\prime_{il}$.
For the Hubbard-Holstein model, this becomes 
\begin{align}\label{Eq:S1}
\nonumber
{W(\sigma^\prime)  \over W(\sigma)} = & \exp\left[ - P_{l,l}^\sigma {  \delta\sigma^2 \over 2} -P^\prime_l \delta\sigma   \right] \times
\\
& \bigg|{ \langle  \psi_L | \exp\left[ i \gamma \delta\sigma (n_{i \uparrow}  ) \right] | \psi_R \rangle \over  \langle \psi_L |  \psi_R \rangle }\bigg|^2 ~,
\end{align}
in which $P_{l,n}^\sigma$ is the propagator obtained after integrating out the phonon degrees of freedom, as derived in  Ref.\,\cite{Karakuzu18}, while $P^\prime_l = \sum\limits_{n} P^\sigma_{l,n} \sigma_{in}$, and $\gamma= \sqrt{U \Delta \tau}$, for the conventional complex auxiliary-field coupled to the  local density $n_{i \uparrow}+n_{i \downarrow}-1$. 
Here we have assumed that the spin-up and spin down components of the weight are factorized, being one the complex conjugate of the other\,\cite{Karakuzu18}, and therefore the left ($\langle \psi_L | $) and right ($| \psi_R \rangle$) wavefunctions are simple Slater determinants of the spin-up electrons. 
Then,   the ratio in Eq.\,\eqref{Eq:S1} becomes
\begin{align}
{W(\sigma^\prime)  \over W(\sigma)} =
&g(\delta \sigma) \exp\left[ - P_{l,l}^\sigma {  \delta\sigma^2 \over 2} -P^\prime_l \delta\sigma   \right] ~, 
\end{align}
with 
\begin{align}
g(\delta \sigma) = \Re\big[B + A  \exp( i \gamma \delta \sigma)\big]~,
\end{align}
where, with a lengthy but  straightforward calculation the constants 
$A$ and $B$ are given by
\begin{eqnarray}
A& =& ( 2 c_\uparrow -2 |c_\uparrow|^2) \nonumber ~,\\
B&=& 1- 2 \Re(c_\uparrow)  +2 |c_\uparrow|^2 \nonumber ~,
\end{eqnarray}
and $c_\uparrow$  is a simple quantity that can be readily evaluated 
during the simulation
\begin{eqnarray}
c_\uparrow &=&  { \langle \psi_L | n_\uparrow | \psi_R \rangle \over 
\langle \psi_L | \psi_R \rangle } .
\end{eqnarray}

In the above equations we have explicitly used that the spin-down component 
in  the weight ratio of Eq.\,\eqref{Eq:S1} 
is the complex conjugate  of the spin-up one, resulting in a  real and positive quantity,  thus implying that $g(\delta \sigma) >0$, as it turns out from the derived expressions  for  $A$  (an explicitly complex 
constant) and $B$ (a real positive one).

Given this, in order to perform the  inverse sampling according to the probability density  $p(\sigma^\prime) \propto  W(\sigma^\prime)$ (see, e.g., Ref.\,\cite{becca17}), one should extract from a random number $0<z<1$, the value of $\delta\sigma$ that satisfies the equality
\begin{align}\label{Eq:z}
\nonumber
z &= \int\limits_{-\infty}^{\delta\sigma} {\rm d}x ~ {1  \over S}  g(x) \exp\left[ - P_{l,l}^\sigma {  x^2 \over 2} -P^\prime_l x   \right]
\\
\nonumber
& = b  \left[1 + {\rm Erf}\bigg({ P^\prime_l+ P^\sigma_{l,l} \delta \sigma \over \sqrt{2 P_{l,l}^\sigma} }\bigg) \right] 
\\
& + \Re\left\{ a \left[1 + {\rm Erf}\bigg({ P^\prime_l -i \gamma + P^\sigma_{l,l} \delta \sigma \over \sqrt{2 P_{l,l}^\sigma} }\bigg) \right] \right\} ~,
\end{align}
with
\begin{align}
\nonumber
b &= {B \over S}  \sqrt{ \pi \over 2  P_{l,l}^\sigma} \exp\left[ { (P^\prime_l)^2 \over 2  P_{l,l}^\sigma } \right]~,
\end{align}
and
\begin{align}
\nonumber
a &= {A \over S}  \sqrt{ \pi \over 2  P_{l,l}^\sigma} \exp\left[ { (P^\prime_l -i \gamma )^2 \over 2  P_{l,l}^\sigma } \right]~.
\end{align}
Here, $S$ is a normalization factor, which is computed as
\begin{equation}
S=\int\limits_{-\infty}^\infty {\rm d}x ~ |g(x)| \exp\left[ - P_{l,l}^\sigma {  x^2 \over 2} -P^\prime_l x   \right] 
\end{equation}
where at half filling yields
\begin{equation}
 S=\sqrt{  2 \pi \over P_{l,l}^\sigma} \left\{  B \exp\left[ {(P^\prime_l)^2 \over  2 P_{l,l}^\sigma}  \right] + \Re( A  \exp\left[ { (P^\prime_l - i \gamma)^2 \over  2  P_{l,l}^\sigma}\right])   \right\} ~.
\end{equation}
The solution of the nonlinear Eq.\,\eqref{Eq:z} is possible with standard 
methods (bisection or Newton methods) and does not affect the 
efficiency of the algorithm for large cluster of of sites.
We have checked in our implementation that this part of the algorithm spends negligible computational resources. 

As a final remark, the complex error function is evaluated using the 
algorithm reported in Ref.\,\onlinecite{Erf}. Notice also that $z=0$ and 1 are not considered in Eq.\,\eqref{Eq:z}, to avoid the sampling of $\delta \sigma=\pm \infty$.
Indeed, these two points can be disregarded without affecting the integral.

\vspace{0.4cm}

\begin{figure}[t]
\centering
\includegraphics[scale=0.26]{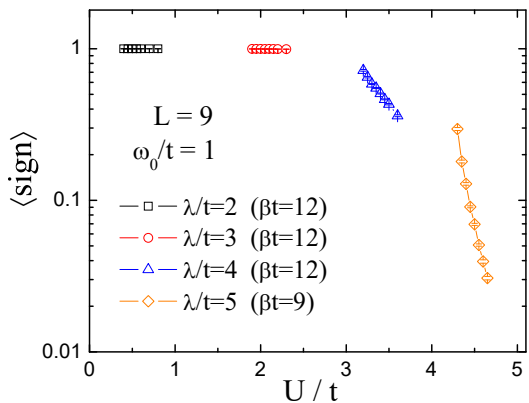}
\caption{Average sign in DQMC simulations of the Hubbard-Holstein model on the honeycomb lattice, for fixed $L=9$ and different interaction strengths.}
\label{fig:sign} 
\end{figure}

\noindent
\underbar{The minus-sign problem in DQMC method}:
The minus-sign problem occurs when the product of determinants in the partition function (used as the statistical weight) assumes negative values~\cite{Loh90,Santos03,gubernatis16}.
It leads to large statistical fluctuations and, consequently, to big error bars, which may prevent the Monte Carlo analysis when the average sign $\langle s \rangle$ is small.
In practice, simulations for $\langle s \rangle \gtrsim 0.1 $ are feasible, in particular when measuring equal-time observables.
For the Hubbard-Holstein model, since the up and down determinants are not equivalent\,\cite{Johnston13}, i.e.~the product is not a positive-definite value, such a problem is expected to appear, with the value of $\langle s \rangle$ depending on the temperature, interaction strengths, and system size.
Therefore, it is important to show the behavior of the average sign in our DQMC simulations, as displayed in Fig.\,\ref{fig:sign}, for fixed $L=9$.
Notice that the average sign is very low for $\lambda/t \gtrsim 5$ (when $\lambda\approx U$).

\vspace{0.1cm}

\pagebreak

\subsection*{Supplementary Tables}

\begin{table}[h]
\begin{ruledtabular}
  \begin{tabular}{cccccc}
     \multicolumn{6}{c}{AFM}
    \\
    \hline
    $\omega_{0}/t$ & & $U/t$ & & $\lambda/t$ & \\
    \hline
 0.5 & & 3.85(2) & &  0.0\footnotemark[1] &  \\
     & & 4.15(4) & &  3.0 & \\
     & & 4.77(4) & &  4.77(4) & \\
    \hline
    
1.0 & & 3.85(2)  & &  0.0\footnotemark[1] & \\
    & & 5.0  & & 3.2(4) & \\
    & & 6.0  & & 5.15(9) & \\
    & & 7.0  & & 6.9(1) & \\
    & & 7.40(4)  & & 7.40(4) & \\
   \hline
$\sqrt{2}$ & & 3.85(2)   & &  0.0\footnotemark[1] &  \\
           & & 5.33(7)   & & 3.0  & \\
           & & 8.0(1)  & & 7.0 & \\
           & & 11.96(5)  & & 11.96(5) & \\
  \end{tabular}
  \end{ruledtabular}
\footnotetext[1]{Reference \cite{Otsuka16}.}
    \caption{Antiferromagnetic (AFM) quantum critical points of the Hubbard-Holstein model on the honeycomb lattice, obtained by projective auxiliary-field quantum Monte Carlo simulations, and presented in Figs.\,1-3.\label{Table1}}
\end{table}

\begin{table}[h]
\begin{ruledtabular}
  \begin{tabular}{cccccc}
     \multicolumn{6}{c}{CDW}
    \\
    \hline
    $\omega_{0}/t$ & & $U/t$ & & $\lambda/t$ & \\
    \hline
    1.0 & & 0.0  & &  1.6(1) & \\
    & & 0.43(6)  & & 2.0 & \\
    & & 1.98(7)  & & 3.0 & \\
    & & 3.23(4)  & & 4.0 & \\
    & & 4.45(5)  & & 5.0\footnotemark[1] & \\
    \hline
    0.25 & & 0.0  & & 1.29(3) & \\
    0.50 & &      & & 1.35(3) & \\
    1.0  & &      & & 1.6(1) & \\
$\sqrt{2}$ & &   & & 1.91(3) & \\
    2.0    & &   & & 2.19(5) & \\
    \hline
    0.50 & &  0.0 & & 1.41(1)\footnotemark[2] & \\
    1.0  & &      & & 1.69(2)\footnotemark[2] & \\
$\sqrt{2}$ & &   & & 1.85(1)\footnotemark[2] & \\
    2.0    & &   & & 2.10(2)\footnotemark[2] & \\
    4.0    & &   & & 2.62(4)\footnotemark[2] & \\
    8.0    & &   & & 3.10(3)\footnotemark[2] & \\
  \end{tabular}
  \end{ruledtabular}
\footnotetext[1]{$U^{\rm cdw}_{c}(6,9)$ for $\beta t=L$ -- see definitions in the main text.}
\footnotetext[2]{$\lambda^{\rm cdw}_{c}(6,9)$ for $\beta t=4L/3$ -- see definitions in the main text.}
\caption{Charge-density wave (CDW) quantum critical points of the Hubbard-Holstein model on the honeycomb lattice, obtained by finite temperature determinant quantum Monte Carlo simulations, and presented in Figs.\,1, 4, and 5. \label{Table2}}
\end{table}

\end{document}